\documentclass[structabstract,printer]{aa} 
\usepackage{graphicx}
\usepackage{longtable}
\usepackage{hyperref}
\usepackage{txfonts}
\usepackage{natbib}
\usepackage{epsfig}
\usepackage{longtable}
\usepackage{times}
\usepackage[english]{babel}
\usepackage{url}

\bibpunct{(}{)}{;}{a}{}{,}

\newcommand{\inte} {\textit{INTEGRAL }}

\newcommand{\sw} {\textit{Swift }}
\newcommand{\swbat} {\textit{Swift}-BAT }
\newcommand{\swxrt} {\textit{Swift}-XRT }

\newcommand{\chandra}{\textit{Chandra }}
\newcommand{\agile}{\textit{AGILE }}

\newcommand {\nh} {N$_{\rm H}$ }

\newcommand {\ferg} {erg cm$^{-2}$ s$^{-1}$ }
\newcommand {\cmmdue} {cm$^{-2}$ }
\newcommand {\hcm} {\hbox {\ifmmode $ atom cm$^{-2}\else atom cm$^{-2}$\fi}}
\begin{document}

\title{The \textit{Swift}-BAT survey reveals the orbital period of three high-mass X-ray binaries}
\author{A.D'A\`i\inst{1}, V.\ La Parola\inst{2}, G.\ Cusumano\inst{2}, A. Segreto \inst{2}, P. Romano\inst{2}, S. Vercellone\inst{2}, 
N. R. Robba\inst{1}}
\institute{ 
Dipartimento di Scienze Fisiche ed Astronomiche, Universit\`a di Palermo, Via Archirafi 36, I-90123, Palermo, Italy
\email{dai@fisica.unipa.it} 
\and
INAF, Istituto di Astrofisica Spaziale e Fisica Cosmica, Via U.\ La Malfa 153, I-90146 Palermo, Italy
}
\date{}
%\pagerange{\pageref{firstpage}--\pageref{lastpage}} \pubyear{2010}

\abstract{}{}{}{}{} 
  \abstract
%context
{}
% aims heading (mandatory)
{A growing number of previously  hidden Galactic X-ray sources are now
  detected  with  recent  surveys  performed  by  the  \inte  and  \sw
  satellites. Most  of these  new sources eluded  past surveys  due to
  their  large local X-ray  extinction and  consequent low  soft X-ray
  flux.  The \swbat performs daily  monitoring of the sky in an energy
  band  (15--150  keV)  which  is  only marginally  affected  by  X-ray
  extinction, thus  allowing for the  search of long  periodicities in
  the  light curve  and  identification  of the  nature  of the  X-ray
  sources.}
% methods heading (mandatory) 
{We  performed a  period search  using  the folding  technique in  the
  \swbat   light    curves   of   three   \inte
  sources:  IGR\,J05007$-$7047, IGR\,J13186$-$6257 and IGR\,J17354$-$3255.    Their
  periodograms   show   significant   peaks   at   $30.77\pm0.01$   d, $19.99\pm0.01$ d
  and $8.448\pm0.002$ d, respectively.  We estimate
  the significance of these features from the $\chi^2$ distribution of
  all the  trials, finding a  probability $\leq$ 1.5$\times$10$^{-4}$   that the
  detections occurred  due to chance.  We complement  our analysis with
  the study of their broadband X-ray emission.}
% results heading (mandatory)
{We  identify  the  periodicities  with  the orbital  periods  of  the
  sources.   The periods  are  typical for  the  wind accretors  X-ray
  binaries and we support  this identification showing that also their
  energy  spectra  are  compatible  with an  X-ray  spectral  emission
  characteristic of high-mass X-ray  binaries. The spectrum of IGR\,J05007$-$704
  that  resides in the  Large Magellanic Cloud, does not show any intrinsic
  local absorption, whereas the spectra of the Galactic sources IGR\,J17354$-$3255 and IGR\,J13186$-$6257 
  may be affected by a local absorber. The folded light curve for IGR\,J13186$-$6257 suggests
  a possible Be companion star.}

%conclusions
{}

%\noindent
%Facility: {\it Swift}
\titlerunning{The orbital period of three HMXBs}
\authorrunning{A. D'A\`i et al.}
\maketitle

%%%%%%%%%%%%%%%%%%%%%%
%%%%%%%%%%%%%%%%%%%%%%
%%%%%%%%%%%%%%%%%%%%%%
%%%%%%%%%%%%%%%%%%%%%%
%%%%%%%%%%%%%%%%%%%%%%
%%%%%%%%%%%%%%%%%%%%%% 

\section{Introduction\label{intro}}
%preamble1

One of the main results of the IBIS/ISGRI telescope \citep{ubertini03}
on board the  \inte satellite \citep{winkler03} is the  detection of a
large number of new hard X-ray sources. At present, \inte has detected
more than  500 sources  ($\sim26\%$ are associated  with extragalactic
sources, $\sim26\%$ with Galactic  sources and $\sim48\%$ whose nature
has not yet been determined)\footnote{An updated list of these sources
  with their main properties and  the relevant references can be found
  in  \url{http://irfu.cea.fr/Sap/IGR-Sources/}}.  Among  the Galactic
sources,  56  are  identified  as  binary systems  with  a  supergiant
companion.  Most  of them,  characterized by persistent  emission, are
likely deeply  embedded in their  stellar wind, as suggested  by their
strong intrinsic  absorption \citep[\nh  $> 10^{23}$ cm$^2$,  see e.g.
][]{walter06}.

An important contribution  to the study of these  new Galactic sources
is provided by the  Burst Alert Telescope (BAT, \citealp{barthelmy05})
on board \sw \citep{gehrels04}, which has been performing a continuous
monitoring of  the sky  in the hard  X-ray energy range  (15--150 keV)
since November 2004.  The telescope, thanks to its large field of view
(1.4  steradians  half coded)  and  its  pointing  strategy, covers  a
fraction between 50\% and 80\% of the sky every day.  This has allowed
the    detection     of    many    of    the     new    \inte    HMXBs
\citep[e.g.][]{cusumano10b}  and  the collection  of  their long  term
light  curves and  of  their  averaged X-ray  spectra.   The long  and
continuous  monitoring  of these  sources  allows  to investigate  the
intrinsic  emission  variability,  to  search for  long  periodicities
(orbital periods) and to discover the presence of eclipse events.  The
role of \swbat  is therefore fundamental to unveil  the nature and the
geometry of these X-ray binary systems.

We are  performing a systematic study  of the light curves  of the new
\inte sgHMXB sources to search for $\geq 0.5$ day periodicities and to
discover  the presence  of eclipse  events allowing  the study  of the
binary system geometry.  In this  paper we present the detailed timing
analysis   and   broadband  (0.2--150   keV)   spectral  analysis   of
IGR\,J05007$-$7047, IGR\,J13186$-$6257, and IGR\,J17354$-$3255.

IGR\,J05007$-$7047 (also known  as IGR\,J05009$-$7044), located in the
Large Magellanic  Cloud (LMC), was  detected with \inte in  the 17--60
keV band  \citep{sazonov05} with a flux  of $1.2\times10^{-11}$ \ferg,
corresponding  to  a luminosity  of  $3.6\times10^{36}$ erg  s$^{-1}$,
assuming a distance of 50 kpc \citep{guainan98}.  A follow-up \chandra
observation allowed  its association  with the relatively  bright blue
($V$=14.8   mag,  $B-V$=-0.01   mag)   star  USNO-B1.0\,0192$-$0057570
\citep{atel572}  of  spectral  type  B2  III  \citep{masetti06}  at  a
redshift consistent  with that  of LMC.  \citet{atel2594}  reported on
the detection in the BAT survey data of a periodic modulation at 30.77
d,  later confirmed by  \citet{atel2597} through  the analysis  of the
optical counterpart light curve.

IGR\,J13186$-$6257 was first reported  in the Third Integral Catalogue
\citep{bird07}  as  an unidentified  hard  X-ray  source. A  follow-up
\swxrt observation better constrained its position and revealed a flat
soft X-ray spectrum \citep{atel1539}.   A short ($\sim$ 5 ks) \chandra
observation found an X-ray source  within the \swxrt error box at (RA,
Dec  [J2000]= 199.604500  deg,  $-$62.970972  deg) \citep{tomsick09}.
The \chandra source, CXOU J131825.0-625815, was identified as the most
likely counterpart of the \inte  source, with a low spurious detection
probability  of $\sim$  5\%.  The  \chandra position  allowed  for the
counterpart   optical    identification   with   the    2MASS   source
J13182505-6258156.  The  spectral shape  of the source  was consistent
with  a highly  absorbed  (1.8 $\times$  10$^{23}$ \cmmdue)  power-law
spectrum.

IGR\,J17354$-$3255  was  discovered during  the  \inte Galactic  Bulge
monitoring program \citep{atel874, kuulkers07}.  The source has also a
possible, highly variable,  $\gamma$-ray counterpart detected with the
\agile  satellite (AGL\,J1734$-$3310,  \citealp{atel2017}).   A \swxrt
follow-up  observation \citep{atel2019} revealed  the presence  of two
sources within the  \inte error circle. Based on  its variability, the
authors  suggested the  source at  (RA, Dec  [J2000]=  263.863167 deg,
$-$32.930250  deg) as  the likely  counterpart  of IGR\,J17354$-$3255.
The spectrum of this source is characterized by strong absorption (\nh
=5$\pm$2 $\times$  10$^{22}$ cm$^{-2}$) and high  flux variability (an
order of magnitude change in  flux within 7 hours). A further \chandra
observation  \citep{tomsick09}  confirmed   its  variable  nature  and
suggested    the    association    with    the   IR    source    2MASS
J17352760$-$3255544.  A flux  modulation in  the \swbat  data,  with a
period of 8.452 days, has been first reported by \citet{atel2596}.

This paper is organized as follows. Section 2 describes the data reduction.
In Sect. 3 and 4 we describe the temporal and spectral analysis and
in Sect. 5 we discuss our results.

\section{Data Reduction\label{data}}

The raw BAT survey data of the first 54 months of the \sw mission were
retrieved          from           the          HEASARC          public
archive\footnote{\url{http://heasarc.gsfc.nasa.gov/docs/archive.html}}
and  processed  with  a  dedicated  software  \citep{segreto10},  that
performs screening, mosaicking and source detection on \swbat data and
produces  spectra  and  light  curves  for  any  given  sky  position.
IGR\,J05007$-$7047,  IGR\,J13186$-$6257  and  IGR\,J17354$-$3255  were
detected in  the \swbat all-sky  map with a significance  maximized in
the  15--50 keV  band of  16.1,  10.0, and  18.9 standard  deviations,
respectively.  Light  curves were extracted in the  15--150 keV energy
range with  the maximum available  time resolution ($\sim 300$  s) and
corrected to the solar system  barycentre (SSB) by using the task {\sc
  earth2sun}.  The  15--150 keV  spectra were obtained  extracting the
source  fluxes from the  sky maps  in sixteen  energy bands,  and were
analyzed using an appropriately rebinned CALDB response matrix.

\swxrt observed  IGR\,J05007$-$7047 between 2010-10-25  and 2010-10-28
(ObsId 31846);  IGR\,J13186$-$6257 was  observed on 2008-05-01  and on
2010-05-25   (ObsId  37071);   IGR\,J17354$-$3255   was  observed   on
2008-03-11  and  2009-04-17  (ObsId  37054).   The  sources  were  all
observed in  Photon Counting mode.  Source and  background spectra and
the  relevant ancillary  files  were created  and  retrieved from  the
\swxrt  product on-line  builder maintained  by the  {\it  Swift} data
center  at the  University of  Leicester \citep{evans09}.   The \swxrt
spectra were rebinned  to have at least 20 counts  per bin, and \swbat
energy  channels were  grouped  to obtain  a  S/N ratio  $\geq$ 3  per
channel.  This  choice allows us  the use of the  $\chi^2$ statistics.
In the following, we report  errors at 90\,\% confidence level, if not
stated otherwise.

Table~\ref{log} reports the log of the \swbat and \swxrt observations
used for the present analysis.

\begin{table*}
\centering
\begin{tabular}{lcccccc}
\hline
\hline
\vspace{0.4cm}
Source         &  $\Delta$T$_{\rm BAT}$ &BAT exposure & XRT ObsId & Obs date &$\Delta$T$_{XRT}$& XRT exposure \\
               &  d                 &d            &           &yyyy-mm-dd& s               & s \\
\hline 
IGR\,J05007$-$7047&   1644.44   & 178.27    & 00031846001 & 2010-10-25 & 28859.7 & 1430.5 \\
               &           &                & 00031846002 & 2010-10-27 & 11894.0 & 971.4 \\
               &           &                & 00031846003 & 2010-10-28 & 35547.4 & 6196.5 \\
IGR\,J13186$-$6257&   1640.80  & 119.73     & 00037071001 & 2008-05-01 &   740   &  739.6 \\
               &           &                & 00037071002 & 2010-05-25 & 29112   & 1531.8 \\ 
IGR\,J17354$-$3255& 1609.18   &  130.30     & 00037054001 & 2008-03-11 & 35254.6 & 4407.7 \\
               &           &                & 00037054002 & 2009-04-17 & 24534.5 & 5285.4 \\
%\vspace{0.4cm}
\hline
\end{tabular}
\caption{Log of \swbat and \swxrt observations used for this work. 
\label{log}}
\end{table*}

%%%%%%%%%%%%%%%%%%%%%%%%%%%%%%%%%%%%%%%%%%%%%%%%%%%%%%%%%
\section{Temporal Analysis \label{temporal}}
%%%%%%%%%%%%%%%%%%%%%%%%%%%%%%%%%%%%%%%%%%%%%%%%%%%%%%%%%

We  analyzed the long  term \swbat  light curve  to search  for intensity
modulations  by applying  a  folding technique  and  searching in  the
0.5--100\,d period range.  The period resolution is
given by P$^{2}/(N  \,\Delta$T$_{\rm BAT})$, where $N=16$ is  the number of
trial profile  phase bins and $\Delta$T$_{\rm BAT}$ is the  data span
length  \citep{buccheri85}.

The  average rate in  each phase  bin was  evaluated by  weighting the
light  curve  rates  by   the  inverse  square  of  the  corresponding
statistical error
\begin{equation}
R_j=\frac{\sum{r_i/er_i^2}}{\sum{1/er_i^2}}  
\end{equation}
where $R_j$  is the average  rate in the  j-th phase bin of  the trial
profile, $r_i$ are  the rate of the light curve  whose phase fall into
the  j-th  phase bin  and  $er_i$  are  the corresponding  statistical
errors.    The  error  on   $R_j$  is   $\left(  \sqrt{\sum{1/er_i^2}}
\right)^{-1}$.  The  weighting procedure was adopted to  deal with the
large span  of $er_i$  and it  is justified because  the BAT  data are
background dominated.

The  significance of  a  feature P$_o$  in  the resulting  periodogram
cannot  be evaluated  using the  $\chi^2$ statistics  because,  if the
source  is   variable  on  time  scales  comparable   to  those  under
investigation, the average $\chi^2$ is far from the value expected for
white  noise  $(N-1)$.   Therefore,   the  significance  of  P$_o$  is
evaluated as follows \citep{laparola10,cusumano10}:  \\ 1. We fit each
periodogram with a  second order polynomial and subtract  the trend to
the    $\chi^2$    distribution,    obtaining   a    new    $\chi^2_C$
distribution.\\  2.  We build  the  histogram  of  the new  $\chi^2_C$
distribution excluding the interval  around P$_o$ and those around any
features deriving  from it (e.g.  periods multiple of P$_o$).\\  3. We
fit  the  tail  of  the  resulting distribution  with  an  exponential
function  and evaluate the  integral of  the best-fit  function beyond
$\chi^2_C(\rm  P_o)$.  This  integral  yields  the  number  of  chance
occurrences due to noise.

\subsection{IGR\,J05007$-$7047}

\begin{figure*}[h]
\begin{center}
%\vspace{-1.5truecm}
\centerline{\psfig{figure=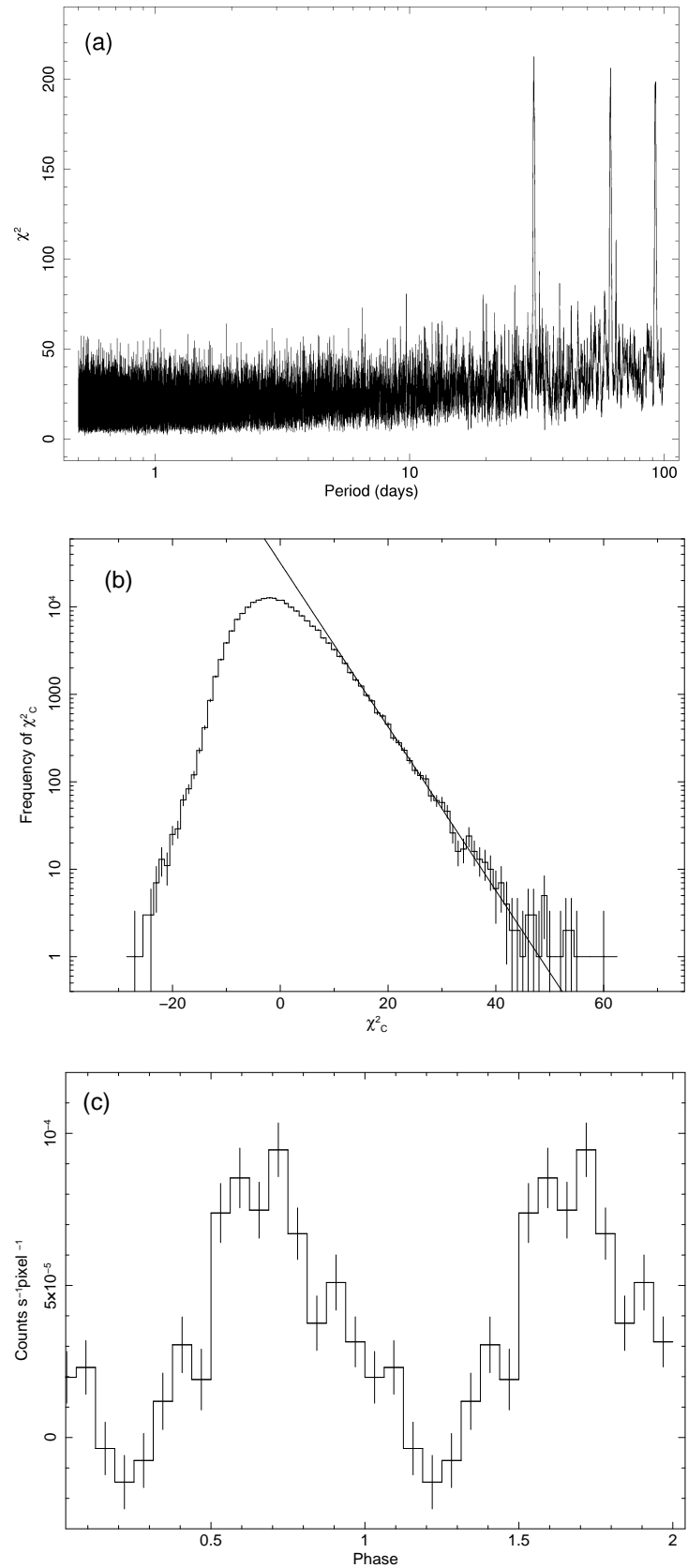,width=7cm,angle=270}}
\vspace{0.5cm}
\centerline{\psfig{figure=fig2.ps,width=7cm,angle=270}}
\vspace{0.5cm}
\centerline{\psfig{figure=fig3.ps,width=7cm,angle=270}}
%\vspace{-2.5truecm}
\caption[]{{\bf (a)}: Periodogram of {\it Swift}-BAT (15--50\,keV) data for 
IGR\,J05007$-$7047.
{\bf (b)}: Distribution of $\chi^2_C$ values excluding the intervals around 
P$_o$ and its multiples. The continuous line is the best fit obtained with an
exponential model  applied to the tail of the distribution ($\chi^2>$ 15).
{\bf (c)}: {\it Swift}-BAT Light curve folded at a period $P=30.77\pm0.01$\,d, 
with 16 phase bins.
                }
		\label{period1} 
\end{center}
\end{figure*}

Figure~\ref{period1}a    shows    the    periodogram   obtained    for
IGR\,J05007$-$7047.  We find significant  evidence for the presence of
a    periodicity     ($\chi^2\sim    210$)    at     a    period    of
$P_{0}=30.77\pm0.03$\,d,  where the period  uncertainty is  the period
resolution  at  P$_o$.  The  other  two  significant  features in  the
periodogram correspond to 2P$_o$  and 3P$_o$.  The value of $\chi^2_C$
(i.e. $\chi^2$  after subtracting the  periodogram trend) at  P$_o$ is
177. Figure~\ref{period1}b shows the $\chi^2_C$ distribution excluding
the values  around P$_o$  and its multiples.  We then fitted  the tail
($\chi^2_C  >15$) of  the resulting  distribution with  an exponential
function and  evaluated the integral  of the best-fit  function beyond
P$_o$.  The  number of chance  occurrences to have  a value of  177 or
higher is $3.9\times 10^{-12}$,  that corresponds to $\sim 7$ standard
deviations    in   Gaussian    statistics.     The   pulsed    profile
(Fig.~\ref{period1}c) folded at  P$_o$ with T$_{\rm epoch}= 54159.753$
MJD,  shows  a roughly  sinusoidal  modulation  with  a minimum  value
consistent  with  zero  intensity.    The  centroid  of  the  minimum,
evaluated by fitting the data around  the dip with a Gaussian model is
at phase  0.23. The  uncertainty depends on  the rough binning  of the
folded light curve and we conservatively assume an error corresponding
to  one  bin phase;  the  dip  phase,  therefore, corresponds  to  MJD
$(54166.8\pm 1.0)\pm  n \times $P$_o$  MJD. The phase coverage  of the
light  curve bins  shows that  this dip  is not  due to  an accidental
under-sampling of the light  curve at this phase. We self-consistently
checked  this  aspect also  for  the  dips  of IGR\,J17354$-$3255  and
IGR\,J13186$-$6257.

The three  {\it Swift}-XRT observations were performed  at the orbital
phase  intervals  of   0.369--0.380,  0.428--0.432  and  0.469--0.482,
respectively.  The 0.2--10 keV light curve shows a persistent emission
with a  variability of a factor  $\sim5$ and an average  count rate of
0.058$\pm$0.004  count s$^{-1}$.  The  source events  selected in  the
three observations  are 77, 37  and 509, respectively. Because  of the
low statistic  content of the  first 2 observations we  restricted the
analysis  to ObsId  00031846003.

\subsection{IGR\,J13186$-$6257}

\begin{figure*}[h]
\begin{center}
%\vspace{-1.5truecm}
\centerline{\psfig{figure=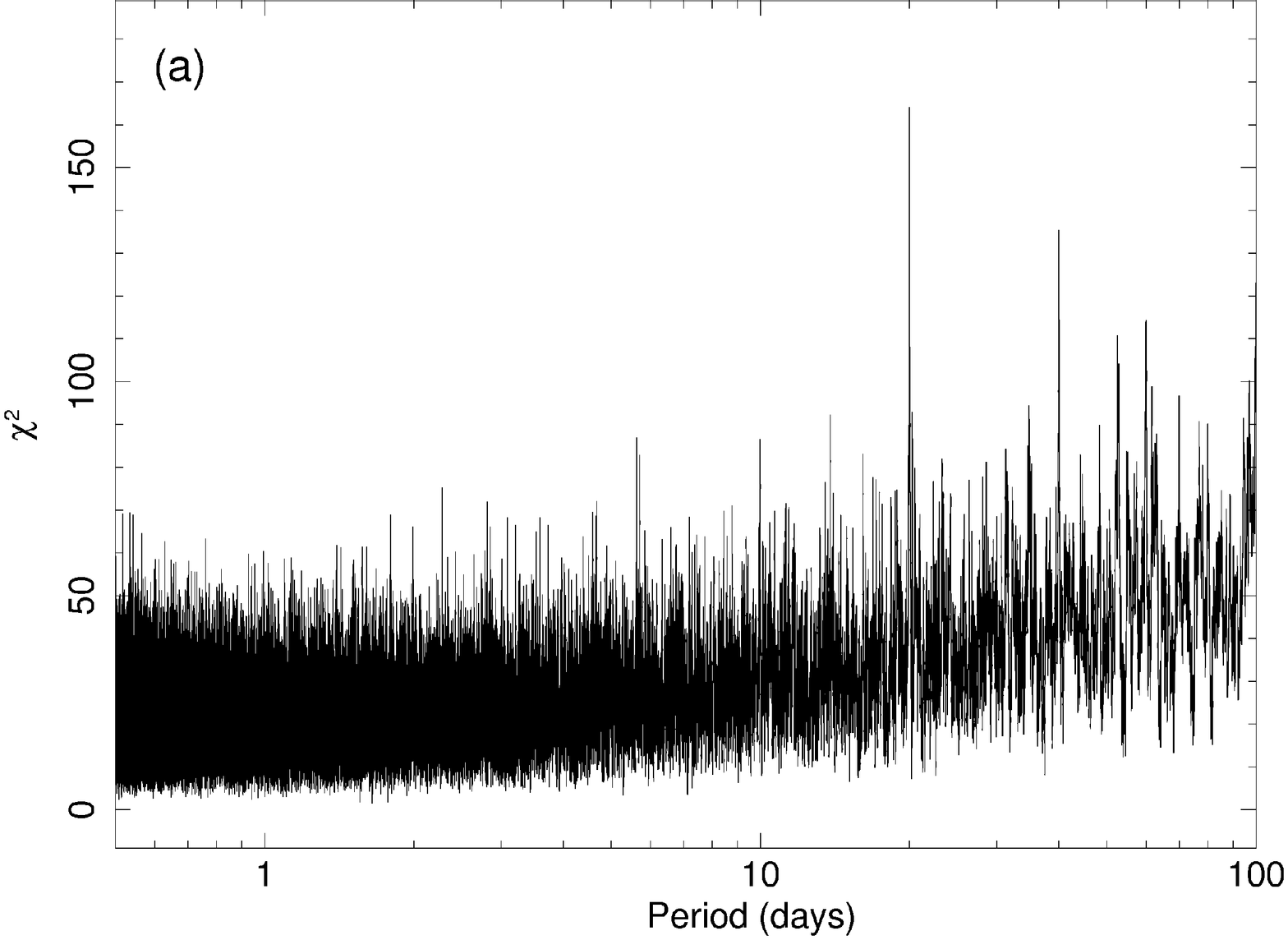,width=7cm,angle=270}}
\vspace{0.5cm}
\centerline{\psfig{figure=fig5.ps,width=7cm,angle=270}}
\vspace{0.5cm}
\centerline{\psfig{figure=fig6.ps,width=7cm,angle=270}}
%\vspace{-2.5truecm}
\caption[]{{\bf (a)}: Periodogram of {\it Swift}-BAT (15--50\,keV) data for 
IGR\,J13186$-$6257.
{\bf (b)}: Distribution of $\chi^2_C$ values excluding the intervals around 
P$_o$ and its multiples. The continuous line is the best fit obtained with an
exponential model  applied to the tail of the distribution ($\chi^2>$ 15).
{\bf (c)}: {\it Swift}-BAT Light curve folded at a period $P=19.99\pm0.01$\,d, 
with 16 phase bins.
                }
		\label{period3} 
\end{center}
\end{figure*}

The  highest feature,  with  a $\chi^2$  value  of $\sim$  164, is  at
P$_o=19.99\pm0.01$  d (Fig.~\ref{period3}a)  .  The  second  and third
highest  peaks in  this periodogram  are found  at multiple  values of
P$_o$ at $\sim$ 40  d and 60 d.  We fit the  periodogram with a second
order  polynomial   and  subtracted   the  trend  from   the  $\chi^2$
distribution, finding  a $\chi^2_C$=126 at P$_o$.   We therefore built
the  histogram of  the  $\chi^2_C$ distribution  (Fig.~\ref{period3}b)
excluding the  interval around  P$_o$ and the  first two  multiples of
P$_o$.  The integral above P$_o$ yields a number of chance occurrences
due to  noise of  $3.2\times10^{-6}$, corresponding to  a significance
for the detected feature  of $\sim4.7$ standard deviations in Gaussian
statistics.

 The pulsed profile (Fig.~\ref{period3}c) folded at P$_o$ with T$_{\rm
   epoch}$=54417.993  MJD, shows a  plateau between  phase 0.9
   and 1.4,  where source emission is dominated  by background,
   whereas the  source becomes clearly detected in  the remaining half
   of the orbit.   We evaluated the time passage  corresponding to the
   folded  light curve  minimum  bin at  phase  0.95 corresponding  to
   T$_{\rm  dip}=(54436.2 \pm 0.6)\pm  n \times  $P$_o$ MJD.   The two
   \swxrt observations  were performed at  orbital phases 0.50  and at
   phase 0.20. The source is clearly detected in the first observation
   with an  averaged count rate  of 0.28$\pm$0.02 count  s$^{-1}$.  In
   the  second observation, performed  two years  later, no  source is
   detected at the same  celestial coordinates.  We estimated an upper
   limit on  the source flux  between 1.2 and 1.5  $\times$ 10$^{-12}$
   \ferg depending on the choice  of the continuum as derived from the
   first observation.  For  spectral analysis we made use  only of the
   data from the first observation.

\subsection{IGR\,J17354$-$3255}

\begin{figure*}[h]
\begin{center}
%\vspace{-1.5truecm}
\centerline{\psfig{figure=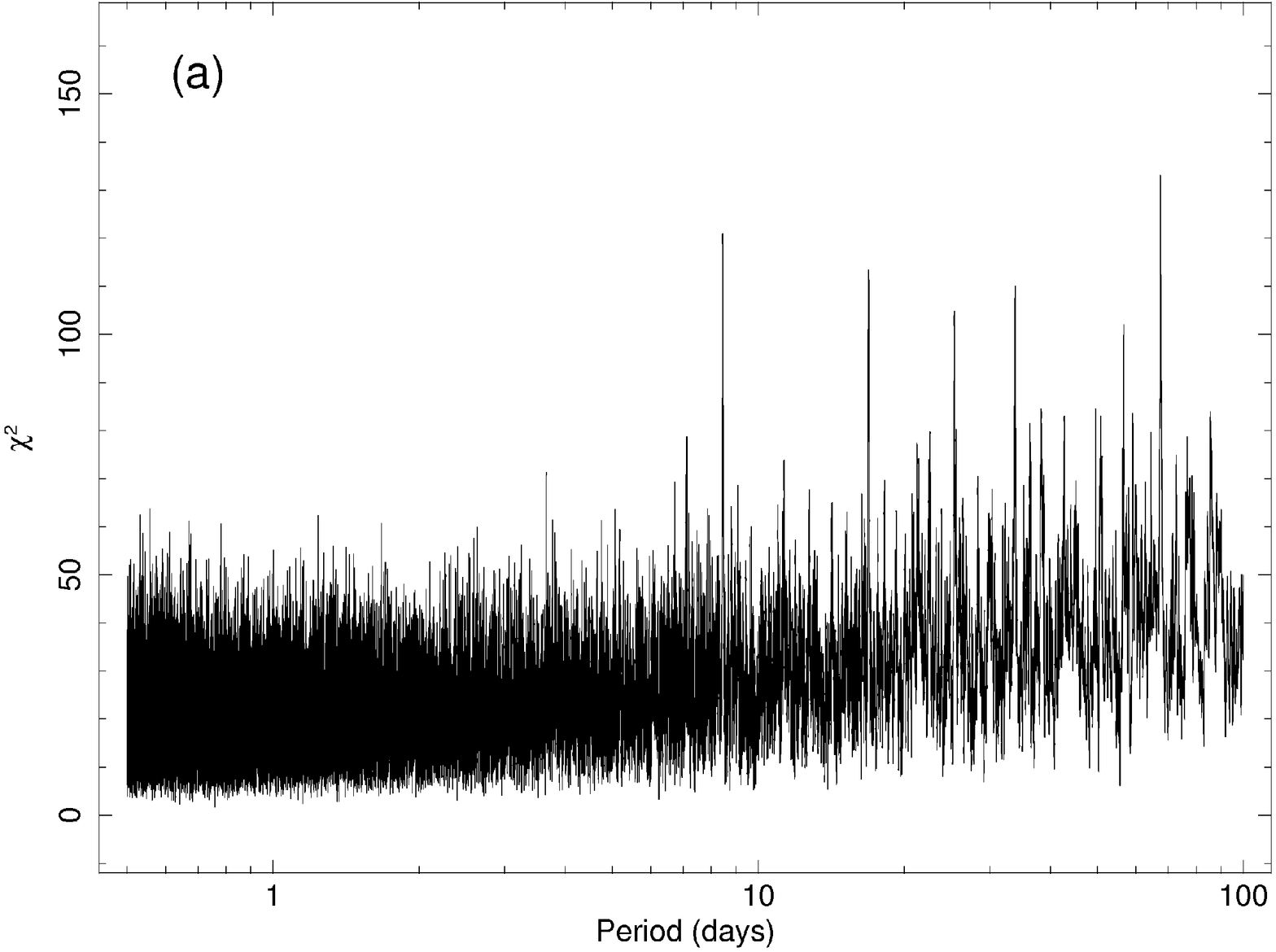,width=7cm,angle=270}}
\vspace{0.5cm}
\centerline{\psfig{figure=fig8.ps,width=7cm,angle=270}}
\vspace{0.5cm}
\centerline{\psfig{figure=fig9.ps,width=7cm,angle=270}}
%\vspace{-2.5truecm}
\caption[]{{\bf (a)}: Periodogram of {\it Swift}-BAT (15--50\,keV) data for 
IGR\,J17354$-$3255.
{\bf (b)}: Distribution of $\chi^2_C$ values excluding the intervals around 
P$_o$ and its multiples. The continuous line is the best fit obtained with an
exponential model  applied to the tail of the distribution ($\chi^2>$ 15).
{\bf (c)}: {\it Swift}-BAT Light curve folded at a period $P=8.452\pm0.002$\,d, 
with 16 phase bins.
                }
		\label{period2} 
\end{center}
\end{figure*}

The       periodogram       obtained      for       IGR\,J17354$-$3255
(Figure~\ref{period2}a) shows significant evidence for the presence of
a    periodicity     ($\chi^2\sim    121$)    at     a    period    of
$P_{0}=8.448\pm0.002$\,d. The  other significant features  that appear
in the  periodogram corresponds to  multiple of P$_o$ (up  to 8P$_o$).
After    correcting   for   the    periodogram   trend,    we   obtain
$\chi^2_C($P$_o)$=94.   Figure~\ref{period2}b   shows  the  $\chi^2_C$
distribution excluding the values around P$_o$ and its multiples.  The
number  of chance  occurrences to  have  a value  of 94  or higher  is
$1.5\times 10^{-4}$  that corresponds to $\sim  4$ standard deviations
in  Gaussian  statistics.   The pulsed  profile  (Fig.~\ref{period2}c)
folded at P$_o$ with T$_{\rm epoch}$=54175.159 MJD, shows a modulation
with a minimum value consistent  with zero intensity.  The centroid of
the  minimum is  at phase  0.78  corresponding to  MJD $(54181.75  \pm
0.26)\pm n \times $P$_o$ MJD.

\begin{figure*}
\begin{center}
\centerline{\psfig{figure=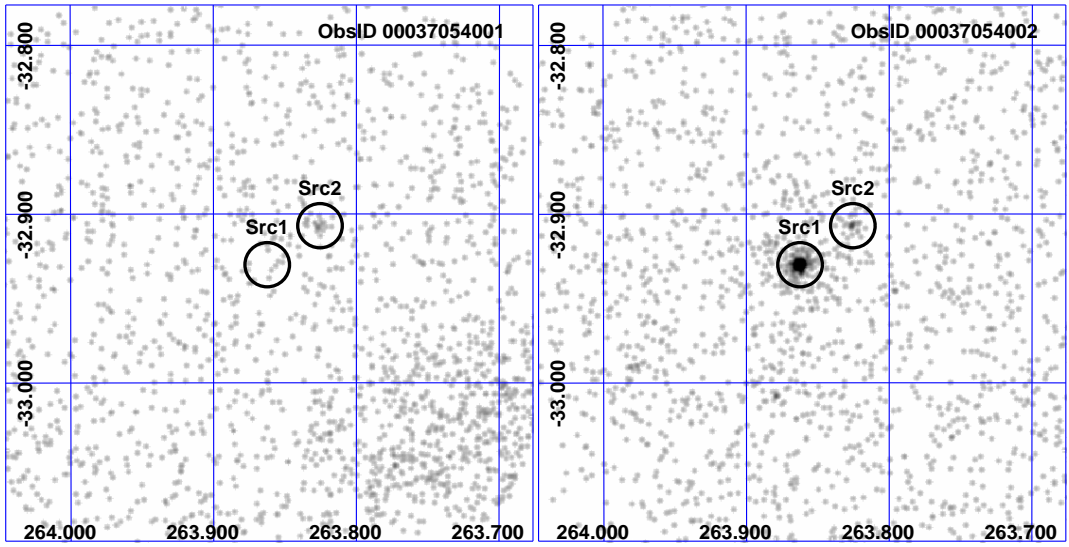,width=14cm,angle=0}
}
\caption[]{Field of view of the two XRT observations of IGR\,J17354$-$3255, with the
position of the two candidate counterparts \citep{atel2019}.
}
\label{xrt} 
\end{center}
\end{figure*}

The  two  \swxrt observations  were  performed  at  the orbital  phase
intervals of 0.719--0.767 and 0.304--0.338, respectively. The phase of
first  observation corresponds to  the minimum  in the  folded profile
(Figure~\ref{period2}c),  while  the phase  of  the  second one  falls
around the  maximum of the  profile.  Figure~\ref{xrt} shows  the {\it
  Swift}-XRT field  of view of  the two observations: the  position of
the two possible counterparts  reported in \citet{atel2019} are marked
as Src1 (RA,  Dec [J2000]= 263.863167 deg, $-$32.930250  deg) and Src2
(RA,  Dec [J2000]=  263.826333 deg,  $-$32.906361 deg).   Src1  is not
detected in  the first observation, with  a 3 $\sigma$  upper limit of
$2.3 \times 10^{-3}$ count  s$^{-1}$.  As this observation corresponds
to a phase  consistent with the minimum of the  BAT folded light curve
we can confidently associate Src1 to IGR J17354$-$3255.  In the second
observation the 0.2--10  keV average count rate of  Src1 is 0.076$\pm$
0.004 count s$^{-1}$,  with a variability of a  factor $\sim 20$.  

%%%%%%%%%%%%%%%%%%%%%%%%%%%%%%%%%%%%%%%%%%%%%%%%%%%%%%%%%
\section{Spectral Analysis \label{spectral}}
%%%%%%%%%%%%%%%%%%%%%%%%%%%%%%%%%%%%%%%%%%%%%%%%%%%%%%%%%

We have  performed a broad band  spectral analysis of the  XRT and BAT
data. The XRT and BAT spectra are not simultaneous, as the BAT spectra
are accumulated  over 54 months.   Therefore, we have checked  for the
presence  of spectral  variability in  the  BAT data  by building  the
hardness  ratio defined as  the ratio  between the  count rate  in the
15--150 keV and in the 15--35  keV energy band, with a time resolution
of  1 day. We  found no  significant variation  of the  hardness ratio
along time for the three sources.  We also verified that the BAT count
rate during the XRT observations  is consistent with the average count
rate in the 54 months.  However, to take into account residual
  flux and intercalibration uncertainties  between the two spectra, we
  included  a multiplicative  constant factor  (C$_{BAT}$) in  all the
  models.  We  kept it frozen to unity  for XRT data and  free to vary
  for BAT  data.  Because of possible correlations  between the values
  of \nh and the indeces  of the power-laws, we built 2D contour-plots
  around the  best-fitting values, always checking  the consistency of
  the error  estimates. 
We  report in Tab.\ref{tab2}  the spectral best-fitting results for the three sources,
using a cut-off power-law model. \swbat fluxes include the C$_{BAT}$ factor
in the calculation.
 
\subsection{IGR\,J05007$-$7047}

\begin{figure*}
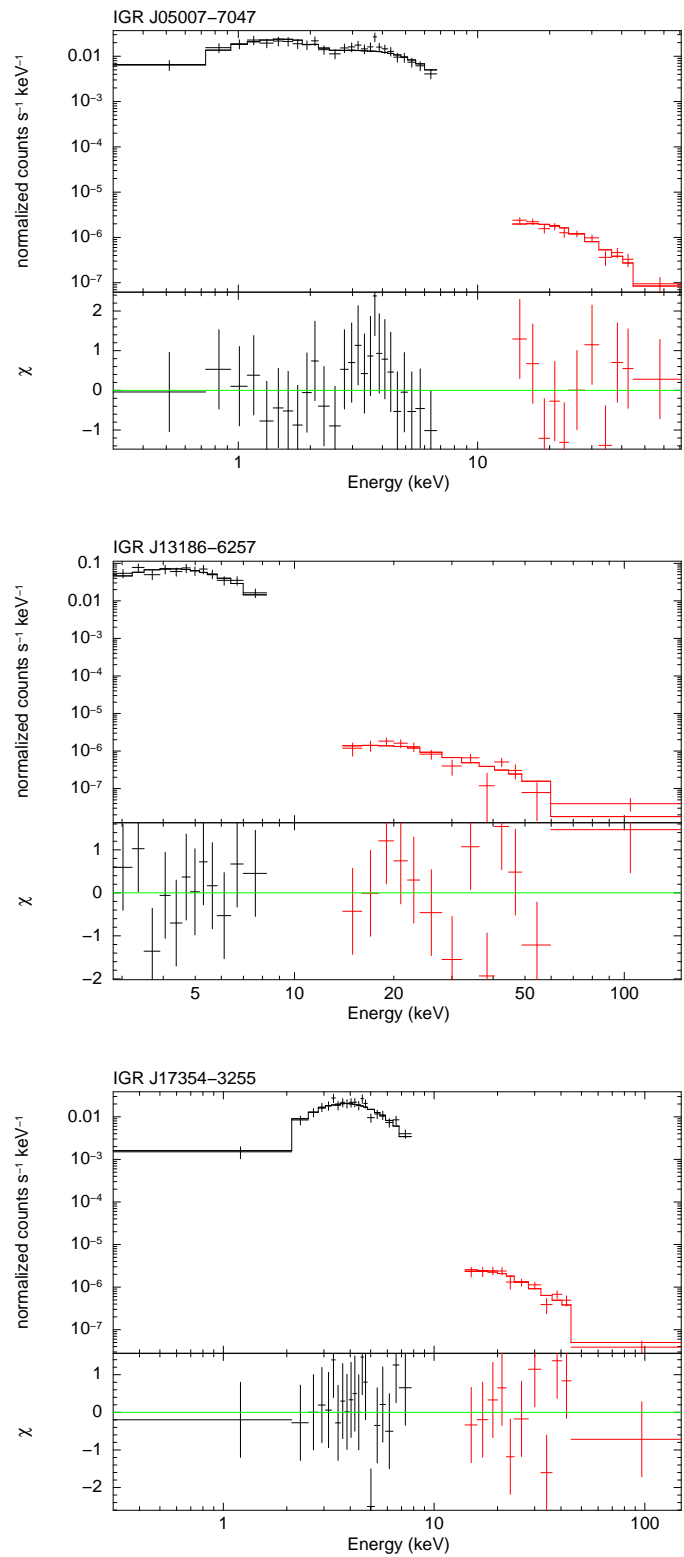

\begin{center}
\centerline{\psfig{figure=fig11.ps, width=6.5cm,angle=270}}
\vspace{0.5cm}
\centerline{\psfig{figure=fig12.ps, width=6.5cm,angle=270}}
\vspace{0.5cm}
\centerline{\psfig{figure=fig13.ps, width=6.5cm,angle=270}}
\caption[]{Combined \swxrt and \swbat spectra, best-fitting model (from Tab.\ref{tab2}) and residuals 
in units of standard deviations of IGR
J05007$-$7047 ({\bf upper panel}), IGR\,J13186-6257  ({\bf middle panel}) and IGR\,J17354$-$3255 ({\bf bottom panel}).}
		\label{spec} 
\end{center}
\end{figure*}

The  combined  XRT-BAT  spectrum  was  first modelled  with  a  simple
absorbed power law.  The resulting  $\chi^2$ (140.2 with 38 degrees of
freedom,  d.o.f.) is  not acceptable,  and the  residuals  suggest the
presence  of a cut-off  at high  energy.  We  have then  included such
cut-off (model {\sc cutoffpl}), obtaining an improved $\chi^2$ of 32.0
for 37 d.o.f., with a cut-off at E$_C=12 \pm 2$ keV and a photon-index
$\Gamma$=0.3$\pm$0.2. The  interstellar absorption is  consistent with
the  expected  value  in  the  direction  of  LMC  \citep{kalberla05}.
Because of  the unambiguous identification of  the optical counterpart
as belonging to the LMC,  we derive an extrapolated 0.2--150 keV X-ray
luminosity   of   $\sim$    9   $\times$   10$^{36}$   erg   s$^{-1}$.
Figure~\ref{spec}a shows the spectrum, best fit model and residuals in
units of standard deviations.

\subsection{IGR\,J13186$-$6257}
An  absorbed  power-law  model   can  satisfactorily  model  the  data
($\chi^2$/d.o.f.=24/21).   For this  model the  \nh  is 17$_{-5}^{+6}$
$\times$ 10$^{22}$ \cmmdue, the power-law photon index is 2.3$\pm$0.3.
The costant  of normalization between  the \swxrt and \swbat  data is,
however,  inconsistent with  unity (being  0.2$_{-0.08}^{+0.14}$).  We
found only marginal evidence for a high-energy cut-off in the spectrum
(E$_{cut}$  $>$ 10  keV), with  a new  $\chi^2$/d.o.f. of  21/20.  The
position of the  high-energy cut-off strongly depends on  the value of
the  \nh and  of  the  photon-index. Choosing  a  reference value  for
E$_{cut}$ at 16 keV, we found an upper limit to the \nh at 11 $\times$
10$^{22}$ \cmmdue and a photon-index constrained in the 0$-$1.0 range.
Using this model the C$_{BAT}$ value is very low (see Tab. \ref{tab2})
,  and  this suggests  strong  variability,  probably  related to  the
orbital  phase.    Taking  into  account   the  time  of   the  \swxrt
observation,  occured  at phase  0.5,  we  extracted a  phase-selected
\swbat spectrum, choosing the  interval 0.5$-$0.8 as representative of
the phase where source  emission is clearly detected above background.
In   this   case,   the   C$_{BAT}$  parameter   assumes   the   value
0.28$_{-0.12}^{+0.22}$,  while  the   spectral  parameters  are  still
consistent with those obtained using the phase-averaged spectrum.  The
sensible  change in the  C$_{BAT}$ parameter  value supports  the idea
that the X-ray  emission may be confined only  in a restricted orbital
phase interval.   We show  in Fig.\ref{spec}b, the  best-fitting model
together  with data  (using  the \swbat  phase-averaged spectrum)  and
residuals in units of $\sigma$.

\subsection{IGR\,J17354$-$3255}

The  fit of  the  combined  XRT-BAT spectrum  with  a simple  absorbed
power-law    model   gives    a   statistically    acceptable   result
($\chi^2$/d.o.f.=29.6/27),  with a  \nh value  of 10.6$_{-1.4}^{+1.7}$
$\times$  10$^{22}$   \cmmdue  and  a   photon-index  of  2.6$\pm$0.3.
However,  fitting the data  with a  cut-off power  law, we  obtained a
significant improvement  of the fit,  with a final  $\chi^2$/d.o.f. of
23.2/26, and  an F-test  probability of $\sim  1\%$ of  obtaining this
improvement  by   chance.   The  cut-off  energy   is  however  poorly
constrained and we obtained only a  lower limit at $\sim$ 18 keV.  The
photon index $\Gamma$ is between 1.0 and 2.0, while the \nh is 7$\pm$2
$\times$ 10$^{22}$ \cmmdue. The expected \nh value in the direction of
the source is considerably  less (1.2-1.6 $\times$ 10$^{22}$ \cmmdue),
and this suggests  that  a fraction  of  the X-ray  absorbing
  medium could be local to the X-ray source.

The source position is only at 5\degr\, from the Galactic Center (GC),
and if  we assume the distance of  the GC as the  possible distance of
the  source (8  kpc), we  obtain an  extrapolated X-ray  luminosity of
$\sim$ 5 $\times$ 10$^{35}$ erg s$^{-1}$.  We show in Fig.\ref{spec}c,
the best-fitting  model together with  data and residuals in  units of
$\sigma$.   

\begin{table*}
\centering
\begin{tabular}{lccc}
\hline\hline
                                    &    IGR\,J05007$-$7047      &   IGR\,J13186$-$6257       & IGR\,J17354$-$3255 \\
\hline                                                                                       
\nh  (10$^{22}$ cm$^{-2}$)           &    $<$ 0.17                & 6$_{-4.5}^{+5}$       & 7$\pm$2  \\
$\Gamma$                            &    0.3$\pm$0.2             & 0.5$\pm$0.5          & 1.7$_{-0.7}^{+0.6}$        \\
E$_{cut}$ (keV)                   &    12$\pm$2                & 16 (\textit{fixed})    & $>$ 18                \\
Flux 0.2-10 keV (10$^{-11}$ \ferg)   &    0.86$_{-0.16}^{+0.13}$   & 5.7$_{-3.0}^{+0.6}$      & 1.1$_{-0.4}^{+0.1}$  \\
Flux 10-100 keV (10$^{-11}$ \ferg)   &    1.9$_{-0.6}^{+0.3}$      & 1.6$\pm$0.4            & 2.6$_{-1.4}^{+0.2}$  \\
$\chi^2$/d.o.f.                     &    25.4/32                 &  21.5/20                 & 23.2/26                \\
C$_{BAT}$                           &    0.8$_{-0.2}^{+0.4}$       & 0.06$_{-0.02}^{+0.14}$  & 1.4$_{-0.7}^{+1.1}$   \\
\hline
\end{tabular}
\caption{Summary of spectral fits using the {\sc cutoffpl} model for the combined \swxrt and \swbat datasets. We report absorbed fluxes
 for the characteristic \swxrt (0.2--10 keV) and \swbat (10--100 keV) 
energy bands. E$_{cut}$ parameter for IGR\,J13186$-$6257 frozen during the fitting.} 
\label{tab2}
\end{table*}

%%%%%%%%%%%%%%%%%%%%%%%%%%%%%%%%%%%%%%%%%%%%%%%%%%%%%%%%%
\section{Discussion and Conclusions \label{disco}}
%%%%%%%%%%%%%%%%%%%%%%%%%%%%%%%%%%%%%%%%%%%%%%%%%%%%%%%%%

In  the  last  census  of  high-mass  X-ray  binaries  in  the  Galaxy
\citep{liu06},  there were 114  HMXBs sources  reported, with  only 50
determined orbital periods. With  an increasing number of new detected
X-ray   sources,    our   view   of   HMXBs    is   rapidly   changing
\citep[see][]{chaty10},  also due  to  the ability  of  the long  term
operating surveys to search for periodic signals.

We are currently undergoing a vast program for the search of periodic
signals exploiting the \swbat surveys \citep{laparola10, cusumano10}.

In this  work, we  have shown  results from the  analysis of  the data
collected  by \swbat during  the first  54 months  of the  {\it Swift}
mission  for  three  still  poorly  understood  \inte  X-ray  sources:
IGR\,J05007$-$7047, IGR\,J13186$-$6257 and IGR J17354$-$3255.

IGR\,J05007$-$7047 was already identified as a possible wind-fed X-ray
system,  thanks   to  the  association  of   the  optical  counterpart
\citep{atel572}.  We  detect it  in the BAT  survey at  a significance
level of $\sim$  16 standard deviations with an  average flux of $\sim
1.3 \times 10^{-11}$ \ferg\ in  the 15--50 keV energy band.  The light
curve reveals a periodicity of $30.77 \pm 0.01$ d that we interpret as
the orbital period of the binary system.  Given the characteristics of
the companion star (spectral  type B2 III, \citealp{masetti06}) we can
apply Kepler's third  law to derive the semi-major  axis of the binary
system.    This    is   given   by    $a^3=P_{\rm   orb}^2\times   \rm
G(M_{\star}+M_{\rm X})/4\pi^2$, where  M$_{\star}$ and M$_{\rm X}$ are
the  masses of  the supergiant  and compact  object,  respectively. We
adopt M$_{\rm X}=1.4$~M$_{\sun}$  and $7 <$M$_{\star} < 12$~M$_{\sun}$
\citep[][for         a        B2         star         of        radius
  $4$~R$_{\sun}<$R$_{\star}<5.6$~R$_{\sun}$]{habets81}.   This  yields
$\rm  84  R_{\sun}<  $$a$$\rm  <  98 R_{\sun}$  ($17  R_{\star}<a<  21
R_{\star}$).   The  folded  profile  shows  a large  dip  at  time  of
$(54166.8\pm 0.6)\pm  nP_o$ MJD with  a count rate consistent  with no
emission.  However,  the width  of  this dip
($\sim20$ per  cent of P$_o$) is not well constrained, constituting a rough
upper limit  too large to be consistent with  the duration of
the eclipse expected for a circular orbit with the inferred semi-major
axis. 

The
spectrum of  IGR\,J05007-7047 is well modelled with  cut-off power law
($\Gamma\sim  0.3$ and  E$_c\sim 12$  keV).   The upper  limit on  the
absorbing column is consistent with  the Galactic value along the line
of sight.  Therefore  this source cannot be classified  as an absorbed
high-mass X-ray binary ($\geq 10^{23}$ cm $^{-2}$).

The  spectral   shape  of  IGR\,J13186$-$6257  could   not  be  firmly
constrained, because  of the weak  statistics of the  \swxrt spectrum.
We could not determine any  clear cut-off in the energy spectrum, with
a  lower limit  at $\sim$  10  keV.  The  periodicity of  19.99 d,  is
typical  for wind/Be  accreting  X-ray systems.  The  folded
  light curve  shows clear source  emission above background  only for
  $\sim$ half of the orbital  period.  Because of the poor constraints
  on the  spectral shape, we are  not able to  univocally classify the
  source.  However, the shape of  the folded light curve, the very low
  value of  the intercalibration  constant between the  pointed \swxrt
  spectrum and  the long-term \swbat  spectrum and the lack  of source
  detection at orbital phase 0.2 suggest a possible Be companion in an
  eccentric  orbit, where  accretion could  take place  only  near the
  peri-astron passage.   The expected equivalent  column density value
  in     the     IGR\,J13186$-$6257     direction     lies     between
  1.24$\times$10$^{22}$       \cmmdue      \citep{kalberla05}      and
  1.59$\times$10$^{22}$ \cmmdue \citep{dickey90}.  These estimates are
  considerably  less than  our  best-fitting values,  both the  simple
  power-law and  for the cut-off  power-law model.  IGR\,J13186$-$6257
  may  be  therefore embedded  in  a  denser  absorbing medium  during
  accretion phases.   Further observations are,  however, necessary to
  confirm this scenario.

IGR\,J17354$-$3255  is  detected  in  the  15--50 keV  BAT  map  at  a
significance level of $\sim19$  standard deviations.  The average flux
in   this  band  is   $\sim  1.56   \times  10^{-11}$   erg  cm$^{-2}$
s$^{-1}$. The timing analysis on the 15--50 BAT light curve unveils an
orbital period of  $8.448\pm0.002$ d, with a significance  of $\sim 4$
standard  deviations, confirming its  nature as  a binary  system. The
detection of the  periodicity in enforced by the  non-detection of the
soft  X-ray counterpart  in  the \swxrt  observation  performed at  an
orbital  phase consistent  with the  minimum of  the  modulation.  The
folded profile  shows a minimum  consistent with zero  intensity.  The
energy spectrum of IGR\,J17354$-$3255  is well modelled by an absorbed
power law, with N$_H\sim 7\times  10^{22}$ cm$^{-2}$ with a cut-off at
$> 20$  keV.  This large local absorption  suggests that X-ray
  emission from the compact object  may interact with the stellar wind
  of  the  companion  star,  whose  spectral type  has  not  yet  been
  determined.

%%%%%%%%%%%%%%%%%%%%%%%%%%%%%%%%%%%%%%%%%%%%%%%%%%%%%%%%%
%%%%%%%%%%%%%%%%%%%%%%%%%%%%%%%%%%%%%%%%%%%%%%%%%%%%%%%%%
%%%%%%%%%%%%%%%%%%%%%%%%%%%%%%%%%%%%%%%%%%%%%%%%%%%%%%%%%
%%%%%%%%%%%%%%%%%%%%%%%%%%%%%%%%%%%%%%%%%%%%%%%%%%%%%%%%%
%%%%%%%%%%%%%%%%%%%%%%%%%%%%%%%%%%%%%%%%%%%%%%%%%%%%%%%%%
%%%%%%%%%%%%%%%%%%%%%%%%%%%%%%%%%%%%%%%%%%%%%%%%%%%%%%%%%

\newpage

\newpage

\newpage

\newpage

\newpage

%%%%%%%%%%%%%%%%%%%%%%%%%%%%%%%%%%%%%%%%%%%%%%%%%%%%%%%%%
%\section*{Acknowledgments}
\begin{acknowledgements}
This work made use of data supplied by the UK Swift Science Data Centre at the 
University of Leicester.
We acknowledge financial contribution from the agreement ASI-INAF I/009/10/0 \\
The authors wish to thank the anonymous referee for comments that helped us to improve the paper.
\end{acknowledgements}
%%%%%%%%%%%%%%%%%%%%%%%%%%%%%%%%%%%%%%%%%%%%%%%%%%%%%%%%%
\bibliographystyle{aa}
\bibliography{refs}

\begin{thebibliography}{30}
\expandafter\ifx\csname natexlab\endcsname\relax\def\natexlab#1{#1}\fi

\bibitem[{{Barthelmy} {et~al.}(2005){Barthelmy}, {Barbier}, {Cummings},
  {Fenimore}, {Gehrels}, {Hullinger}, {Krimm}, {Markwardt}, {Palmer},
  {Parsons}, {Sato}, {Suzuki}, {Takahashi}, {Tashiro}, \&
  {Tueller}}]{barthelmy05}
{Barthelmy}, S.~D., {Barbier}, L.~M., {Cummings}, J.~R., {et~al.} 2005, \ssr,
  120, 143

\bibitem[{{Bird} {et~al.}(2007){Bird}, {Malizia}, {Bazzano}, {Barlow},
  {Bassani}, {Hill}, {B{\'e}langer}, {Capitanio}, {Clark}, {Dean}, {Fiocchi},
  {G{\"o}tz}, {Lebrun}, {Molina}, {Produit}, {Renaud}, {Sguera}, {Stephen},
  {Terrier}, {Ubertini}, {Walter}, {Winkler}, \& {Zurita}}]{bird07}
{Bird}, A.~J., {Malizia}, A., {Bazzano}, A., {et~al.} 2007, \apjs, 170, 175

\bibitem[{{Buccheri} \& {Sacco}(1985)}]{buccheri85}
{Buccheri}, R. \& {Sacco}, B. 1985, in Data Analysis in Astronomy, ed. {V.~di
  Gesu, L.~Scarsi, P.~Crane, J.~H.~Friedman, \& S.~Levialdi }, 15

\bibitem[{{Bulgarelli} {et~al.}(2009){Bulgarelli}, {Gianotti}, {Trifoglio},
  {Striani}, {Tavani}, {Sabatini}, {Vercellone}, {Feroci}, {Lazzarotto}, {Del
  Monte}, {Pittori}, {Verrecchia}, {Pellizzoni}, {Pilia}, {Chen}, {Giuliani},
  {D'Ammando}, {Piano}, {Pucella}, {Vittorini}, {Costa}, {Donnarumma},
  {Pacciani}, {Soffitta}, {Evangelista}, {Lapshov}, {Rapisarda}, {Argan},
  {Trois}, {de Paris}, {Marisaldi}, {Di Cocco}, {Labanti}, {Fuschino}, {Galli},
  {Caraveo}, {Mereghetti}, {Perotti}, {Fiorini}, {Zambra}, {Barbiellini},
  {Longo}, {Moretti}, {Vallazza}, {Picozza}, {Morselli}, {Prest}, {Lipari},
  {Zanello}, {Cattaneo}, {Santolamazza}, {Colafrancesco}, {Giommi}, \&
  {Salotti}}]{atel2017}
{Bulgarelli}, A., {Gianotti}, F., {Trifoglio}, M., {et~al.} 2009, The
  Astronomer's Telegram, 2017, 1

\bibitem[{{Chaty} {et~al.}(2010){Chaty}, {Zurita Heras}, \&
  {Bodaghee}}]{chaty10}
{Chaty}, S., {Zurita Heras}, J.~A., \& {Bodaghee}, A. 2010, ArXiv e-prints

\bibitem[{{Coe} {et~al.}(2010){Coe}, {Townsend}, \& {Udalski}}]{atel2597}
{Coe}, M.~J., {Townsend}, L.~J., \& {Udalski}, A. 2010, The Astronomer's
  Telegram, 2597, 1

\bibitem[{{Cusumano} {et~al.}(2010{\natexlab{a}}){Cusumano}, {La Parola},
  {Romano}, {Segreto}, {Vercellone}, \& {Chincarini}}]{cusumano10}
{Cusumano}, G., {La Parola}, V., {Romano}, P., {et~al.} 2010{\natexlab{a}},
  \mnras, 406, L16

\bibitem[{{Cusumano} {et~al.}(2010{\natexlab{b}}){Cusumano}, {La Parola},
  {Segreto}, {Mangano}, {Ferrigno}, {Maselli}, {Romano}, {Mineo}, {Sbarufatti},
  {Campana}, {Chincarini}, {Giommi}, {Masetti}, {Moretti}, \&
  {Tagliaferri}}]{cusumano10b}
{Cusumano}, G., {La Parola}, V., {Segreto}, A., {et~al.} 2010{\natexlab{b}},
  \aap, 510, A48

\bibitem[{{D'A\`i} {et~al.}(2010){D'A\`i}, {Cusumano}, {La Parola}, \&
  {Segreto}}]{atel2596}
{D'A\`i}, A., {Cusumano}, G., {La Parola}, V., \& {Segreto}, A. 2010, The
  Astronomer's Telegram, 2596, 1

\bibitem[{{Dickey} \& {Lockman}(1990)}]{dickey90}
{Dickey}, J.~M. \& {Lockman}, F.~J. 1990, \araa, 28, 215

\bibitem[{{Evans} {et~al.}(2009){Evans}, {Beardmore}, {Page}, {Osborne},
  {O'Brien}, {Willingale}, {Starling}, {Burrows}, {Godet}, {Vetere}, {Racusin},
  {Goad}, {Wiersema}, {Angelini}, {Capalbi}, {Chincarini}, {Gehrels}, {Kennea},
  {Margutti}, {Morris}, {Mountford}, {Pagani}, {Perri}, {Romano}, \&
  {Tanvir}}]{evans09}
{Evans}, P.~A., {Beardmore}, A.~P., {Page}, K.~L., {et~al.} 2009, \mnras, 397,
  1177

\bibitem[{{Gehrels} {et~al.}(2004){Gehrels}, {Chincarini}, {Giommi}, {Mason},
  {Nousek}, {Wells}, {White}, {Barthelmy}, {Burrows}, {Cominsky}, {Hurley},
  {Marshall}, {M{\'e}sz{\'a}ros}, {Roming}, {Angelini}, {Barbier}, {Belloni},
  {Campana}, {Caraveo}, {Chester}, {Citterio}, {Cline}, {Cropper}, {Cummings},
  {Dean}, {Feigelson}, {Fenimore}, {Frail}, {Fruchter}, {Garmire}, {Gendreau},
  {Ghisellini}, {Greiner}, {Hill}, {Hunsberger}, {Krimm}, {Kulkarni}, {Kumar},
  {Lebrun}, {Lloyd-Ronning}, {Markwardt}, {Mattson}, {Mushotzky}, {Norris},
  {Osborne}, {Paczynski}, {Palmer}, {Park}, {Parsons}, {Paul}, {Rees},
  {Reynolds}, {Rhoads}, {Sasseen}, {Schaefer}, {Short}, {Smale}, {Smith},
  {Stella}, {Tagliaferri}, {Takahashi}, {Tashiro}, {Townsley}, {Tueller},
  {Turner}, {Vietri}, {Voges}, {Ward}, {Willingale}, {Zerbi}, \&
  {Zhang}}]{gehrels04}
{Gehrels}, N., {Chincarini}, G., {Giommi}, P., {et~al.} 2004, \apj, 611, 1005

\bibitem[{{Guinan} {et~al.}(1998){Guinan}, {Fitzpatrick}, {Dewarf}, {Maloney},
  {Maurone}, {Ribas}, {Pritchard}, {Bradstreet}, \& {Gim{\'e}nez}}]{guainan98}
{Guinan}, E.~F., {Fitzpatrick}, E.~L., {Dewarf}, L.~E., {et~al.} 1998, \apjl,
  509, L21

\bibitem[{{Habets} \& {Heintze}(1981)}]{habets81}
{Habets}, G.~M.~H.~J. \& {Heintze}, J.~R.~W. 1981, \aaps, 46, 193

\bibitem[{{Halpern}(2005)}]{atel572}
{Halpern}, J.~P. 2005, The Astronomer's Telegram, 572, 1

\bibitem[{{Kalberla} {et~al.}(2005){Kalberla}, {Burton}, {Hartmann}, {Arnal},
  {Bajaja}, {Morras}, \& {P{\"o}ppel}}]{kalberla05}
{Kalberla}, P.~M.~W., {Burton}, W.~B., {Hartmann}, D., {et~al.} 2005, \aap,
  440, 775

\bibitem[{{Kuulkers} {et~al.}(2006){Kuulkers}, {Shaw}, {Paizis}, {Gros},
  {Chenevez}, {Sanchez-Fernandez}, {Brandt}, {Courvoisier}, {Garau}, {Ebisawa},
  {Kretschmar}, {Markwardt}, {Mowlavi}, {Oosterbroek}, {Orr}, {Oneca}, \&
  {Wijnands}}]{atel874}
{Kuulkers}, E., {Shaw}, S., {Paizis}, A., {et~al.} 2006, The Astronomer's
  Telegram, 874, 1

\bibitem[{{Kuulkers} {et~al.}(2007){Kuulkers}, {Shaw}, {Paizis}, {Chenevez},
  {Brandt}, {Courvoisier}, {Domingo}, {Ebisawa}, {Kretschmar}, {Markwardt},
  {Mowlavi}, {Oosterbroek}, {Orr}, {R{\'{\i}}squez}, {Sanchez-Fernandez}, \&
  {Wijnands}}]{kuulkers07}
{Kuulkers}, E., {Shaw}, S.~E., {Paizis}, A., {et~al.} 2007, \aap, 466, 595

\bibitem[{{La Parola} {et~al.}(2010{\natexlab{a}}){La Parola}, {Cusumano},
  {Romano}, {Segreto}, {Vercellone}, \& {Chincarini}}]{laparola10}
{La Parola}, V., {Cusumano}, G., {Romano}, P., {et~al.} 2010{\natexlab{a}},
  \mnras, 405, L66

\bibitem[{{La Parola} {et~al.}(2010{\natexlab{b}}){La Parola}, {Cusumano},
  {Segreto}, {Romano}, {Vercellone}, \& {D'Ai'}}]{atel2594}
{La Parola}, V., {Cusumano}, G., {Segreto}, A., {et~al.} 2010{\natexlab{b}},
  The Astronomer's Telegram, 2594, 1

\bibitem[{{Landi} {et~al.}(2008){Landi}, {Masetti}, {Malizia}, {Del Santo},
  {Tarana}, {Bird}, {Dean}, {Caraveo}, \& {Senziani}}]{atel1539}
{Landi}, R., {Masetti}, N., {Malizia}, A., {et~al.} 2008, The Astronomer's
  Telegram, 1539, 1

\bibitem[{{Liu} {et~al.}(2006){Liu}, {van Paradijs}, \& {van den
  Heuvel}}]{liu06}
{Liu}, Q.~Z., {van Paradijs}, J., \& {van den Heuvel}, E.~P.~J. 2006, \aap,
  455, 1165

\bibitem[{{Masetti} {et~al.}(2006){Masetti}, {Morelli}, {Palazzi}, {Galaz},
  {Bassani}, {Bazzano}, {Bird}, {Dean}, {Israel}, {Landi}, {Malizia},
  {Minniti}, {Schiavone}, {Stephen}, {Ubertini}, \& {Walter}}]{masetti06}
{Masetti}, N., {Morelli}, L., {Palazzi}, E., {et~al.} 2006, \aap, 459, 21

\bibitem[{{Sazonov} {et~al.}(2005){Sazonov}, {Churazov}, {Revnivtsev},
  {Vikhlinin}, \& {Sunyaev}}]{sazonov05}
{Sazonov}, S., {Churazov}, E., {Revnivtsev}, M., {Vikhlinin}, A., \& {Sunyaev},
  R. 2005, \aap, 444, L37

\bibitem[{{Segreto} {et~al.}(2010){Segreto}, {Cusumano}, {Ferrigno}, {La
  Parola}, {Mangano}, {Mineo}, \& {Romano}}]{segreto10}
{Segreto}, A., {Cusumano}, G., {Ferrigno}, C., {et~al.} 2010, \aap, 510, A47

\bibitem[{{Tomsick} {et~al.}(2009){Tomsick}, {Chaty}, {Rodriguez}, {Walter}, \&
  {Kaaret}}]{tomsick09}
{Tomsick}, J.~A., {Chaty}, S., {Rodriguez}, J., {Walter}, R., \& {Kaaret}, P.
  2009, \apj, 701, 811

\bibitem[{{Ubertini} {et~al.}(2003){Ubertini}, {Lebrun}, {Di Cocco}, {Bazzano},
  {Bird}, {Broenstad}, {Goldwurm}, {La Rosa}, {Labanti}, {Laurent}, {Mirabel},
  {Quadrini}, {Ramsey}, {Reglero}, {Sabau}, {Sacco}, {Staubert}, {Vigroux},
  {Weisskopf}, \& {Zdziarski}}]{ubertini03}
{Ubertini}, P., {Lebrun}, F., {Di Cocco}, G., {et~al.} 2003, \aap, 411, L131

\bibitem[{{Vercellone} {et~al.}(2009){Vercellone}, {D'Ammando}, {Striani},
  {Tavani}, {Sabatini}, {Bulgarelli}, {Gianotti}, {Trifoglio}, {Feroci},
  {Lazzarotto}, {Del Monte}, {Pittori}, {Verrecchia}, {Pellizzoni}, {Pilia},
  {Chen}, {Giuliani}, {Piano}, {Pucella}, {Vittorini}, {Costa}, {Donnarumma},
  {Pacciani}, {Soffitta}, {Evangelista}, {Lapshov}, {Rapisarda}, {Argan},
  {Trois}, {de Paris}, {Marisaldi}, {Di Cocco}, {Labanti}, {Fuschino}, {Galli},
  {Caraveo}, {Mereghetti}, {Perotti}, {Fiorini}, {Zambra}, {Barbiellini},
  {Longo}, {Moretti}, {Vallazza}, {Picozza}, {Morselli}, {Prest}, {Lipari},
  {Zanello}, {Cattaneo}, {Rappoldi}, {Santolamazza}, {Colafrancesco}, {Giommi},
  {Salotti}, {Romano}, {Burrows}, \& {Gehrels}}]{atel2019}
{Vercellone}, S., {D'Ammando}, F., {Striani}, E., {et~al.} 2009, The
  Astronomer's Telegram, 2019, 1

\bibitem[{{Walter} {et~al.}(2006){Walter}, {Zurita Heras}, {Bassani},
  {Bazzano}, {Bodaghee}, {Dean}, {Dubath}, {Parmar}, {Renaud}, \&
  {Ubertini}}]{walter06}
{Walter}, R., {Zurita Heras}, J., {Bassani}, L., {et~al.} 2006, \aap, 453, 133

\bibitem[{{Winkler} {et~al.}(2003){Winkler}, {Courvoisier}, {Di Cocco},
  {Gehrels}, {Gim{\'e}nez}, {Grebenev}, {Hermsen}, {Mas-Hesse}, {Lebrun},
  {Lund}, {Palumbo}, {Paul}, {Roques}, {Schnopper}, {Sch{\"o}nfelder},
  {Sunyaev}, {Teegarden}, {Ubertini}, {Vedrenne}, \& {Dean}}]{winkler03}
{Winkler}, C., {Courvoisier}, T., {Di Cocco}, G., {et~al.} 2003, \aap, 411, L1

\end{thebibliography}
\end{document}